\documentclass[showpacs,twocolumn,amsmath,superscriptaddress]{revtex4-1}
\usepackage[dvipdfmx]{graphicx}
\usepackage{color,amsmath,lmodern,bm,array}

\newcommand{\bk}{{\bm k}}
\newcommand{\bx}{{\bm x}}

\begin{document}

\preprint{UPt3}

\title{Exotic multi-gap structure in UPt$_3$ unveiled by the first-priniciples analysis}%

\author{Takuya Nomoto}%
\email{nomoto.takuya@scphys.kyoto-u.ac.jp}
\affiliation{Department of Physics, Kyoto University, Kyoto, 606-8502, Japan}%
\author{Hiroaki Ikeda}%
\affiliation{Department of Physics, Ritsumeikan University, Kusatsu, 525-8577, Japan}%
\date{\today}%
\begin{abstract}
A heavy-fermion superconductor UPt$_3$ is a unique spin-triplet superconductor with multiple superconducting phases. Here we provide the first report on the first-principles analysis of the microscopic superconducting gap structure. We find that the promising gap structure is an unprecedented $E_{2u}$ state, which is completely different from the previous phenomenological $E_{2u}$ models. Our obtained $E_{2u}$ state has in-plane twofold vertical line nodes on small Fermi surfaces and point nodes with linear dispersion on a large Fermi surface. These peculiar features cannot be explained in the conventional spin $1/2$ representation, but is described by the group-theoretical representation of the Cooper pairs in the total angular momentum $j=5/2$ space. Our findings shed new light on the long-standing problems in the superconductivity of UPt$_3$.

\end{abstract}


\maketitle

Identifying the pairing state and the pairing mechanism is one of the most interesting and important issues in the field of unconventional superconductivity. In particular,  a spin-triplet type of pairing state attracts much attention, since there are few examples except for the superfluid helium 3. In the strongly correlated electron systems, the heavy-fermion superconductor UPt$_3$ is one of the rare candidates for spin-triplet superconductors~\cite{Sauls,Joynt}. The most impressive feature of this material is the multiple superconducting phase diagram. At zero magnetic field, there appears the superconducting double transition into the $A$ phase at the upper critical temperature $T_c^+\sim 540$mK, and then into the $B$ phase at the lower $T_c^-\sim 490$mK \![\onlinecite{Fisher}]. Moreover, the $C$ phase appears at high field and low temperature in the $H-T$ phase diagram~\cite{Bruls,Adenwalla}. In each phase, nodal quasiparticle excitations have been observed~\cite{Shivaram2,Brison,Lussier,Suderow}, and also the time-reversal symmetry breaking has been reported in $B$ phase~\cite{TRS1,TRS2,TRS3}. In spite of these prominent features, the superconducting gap structure still remains to be solved. Many scenarios have been proposed based on the phenomenological approach so far~\cite{Blount2,Agterberg,Zhitomirsky,Krotkov}. Among them, the most promising gap symmetry has been widely believed to be $E_{2u}$ models~\cite{Choi2,Sauls,Sauls2,Graf}.  However, recent measurement of the field-angle resolved thermal transport has detected in-plane twofold oscillations in the $C$ phase~\cite{YMachida}. This result is inconsistent with the proposed $E_{2u}$ models, because in the group-theoretical argument, it is believed that the $E_{2u}$ models do not have such in-plane twofold symmetry. Such twofold symmetry seems to be rather compatible with the $E_{1u}$ models proposed in Refs.~[\onlinecite{Machida1,Machida2,Tsutsumi}]. This is also supported by the following observations. A small residual thermal conductivity~\cite{Izawa} suggests the presence of point nodes with linear dispersion in the $E_{1u}$ models. The Josephson effect~\cite{Gouchi} with $s$-wave superconductor is compatible with $E_{1u}$ planar states. Thus, recently, the $E_{1u}$ models~\cite{Machida1,Machida2,Tsutsumi} have been revisited. This strongly promoted the field-angle resolved specific heat measurement. However, the complimentary measurements have not detected any signature of in-plane symmetry breaking in any phases~\cite{Kittaka}. Although this seems to contradict the result in the thermal conductivity, it is expected to be explained by considering the multi-band nature of UPt$_3$. If the twofold vertical line nodes are located on the Fermi surface (FS) with a light band mass, then the twofold oscillations will be more remarkable in the thermal conductivity than the specific heat measurement. In order to clarify how reasonable such plausible story is, the microscopic analysis of superconductivity including the electronic structure in UPt$_3$ is worth consideration~\cite{Norman87,Putikka,Norman94}. In this regard, recent progress on the first-principles theoretical approach allows us to investigate the gap structure microscopically even in the complicated band structure like the heavy-fermion compounds~\cite{Ikeda1,Nomoto,Ikeda2}. 

In this letter, we provide the first report on a microscopic theory of superconductivity in UPt$_3$ based on the first-principles approach. Generally, it is difficult to exactly evaluate the effect of strong electron correlation in $f$-electron materials. Instead, we study probable candidates of gap functions based on the Fermi-liquid picture as a first step to understand unconventional superconductivity in UPt$_3$~[\onlinecite{memo}]. We find that the promising gap structure is an unprecedented $E_{2u}$ pairing state,  which is supported by the $j=5/2$ representation of Cooper pairs, instead of conventional pseudo-spin representations. Its nodal structure is completely different on each FS; the point nodes with linear dispersion in the large hole FS, and the twofold vertical line nodes in small electron FSs. These features are not expected in the well-known phenomenological $E_{2u}$ model. The low-energy nodal excitations are similar to those in the $E_{1u}$ model rather than the previous $E_{2u}$ model. The peculiar properties can give a comprehensive explanation for the above-mentioned experimental observations, including the seemingly inconsistent result between the thermal conductivity and specific-heat measurement. Thus, the exotic $E_{2u}$ gap structure is the most promising pairing state in the superconductivity of UPt$_3$.

\paragraph*{Fermi surface and model Hamiltonian~---}
\begin{figure}[t]
\centering
\includegraphics[width=8.0cm]{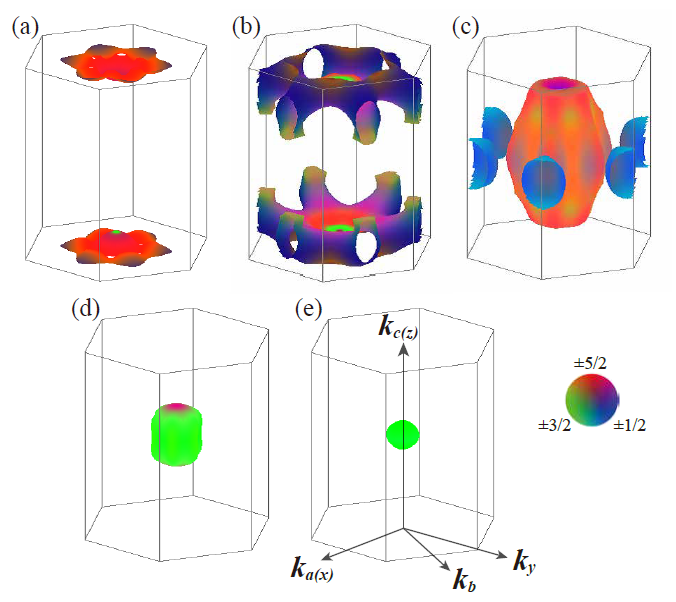}
\caption{Orbital-resolved Fermi surfaces in our tight-binding model $H_0$, obtained by the first-principles calculations. The colors correspond to the weight of $j_z$ component in the total angular momentum $j=5/2$ space. In the text, (a)-(e) are referred to as band$1$-$5$, respectively.}
\end{figure}

In studying the superconductivity of UPt$_3$, the itinerant 5f model is considered to be a good starting point, since the Fermi surface in the first-principles calculations has been partially supported by the de Haas van Alphen measurements~\cite{Taillefer,Kimura}. Following the previous studies~\cite{Ikeda1,Ikeda2}, we here figure out the magnetic fluctuations in UPt$_3$, based on the first-principles theoretical approach.

First of all, using the {\tt WIEN2k} package~\cite{Wien2k}, we calculate the electronic structure of UPt$_3$, and then construct an effective tight-binding model~\cite{supple} in the Wannier bases using the {\tt wien2wannier} interface~\cite{Kunes} and the {\tt wannier90} code~\cite{Wannier90}. Here we employ the space group $P6_3/mmc$, which holds the in-plane six-fold rotational symmetry. Note that the so-called symmetry breaking term is not included~\cite{Midgley,crystal}. Our model Hamiltonian is composed of 120 Wannier bases, containing U(5f), U(6d), Pt(5d), Pt(6s) orbitals and spin degrees of freedom. These bases are transformed into the bases of the total angular momentum $j$. In this case, due to the moderate spin-orbit coupling, the orbital components of the bands crossing the Fermi level are dominated by the $j=5/2$ multiplet of U(5f) orbitals, and the $j=7/2$ multiplet is located at much higher position.

The obtained FS is illustrated in Fig.1. Colors on the FS, red, green, and blue, correspond to each weight of $j_z=\pm 5/2$, $\pm 3/2$ and $\pm 1/2$ components, respectively. The FS topology is well consistent with the previous studies~\cite{Joynt,Norman2,band}. The FSs of Figs.1(b) and (c) have a large contribution to the density of states (DOS) at the Fermi level. Here we realize that each FS possesses relatively separated orbital components, especially, the small FSs in Figs.1(d) and (e) roughly involve only $j_z=\pm 3/2$ component. This characteristic feature is the key to the emergence of the unprecedented $E_{2u}$ gap structure as discuss below.

\paragraph*{Magnetic fluctuations~---}
Next we study the magnetic fluctuations in the model Hamiltonian, including the on-site Hubbard-type repulsions, $U, U', J, J'$ between 5f electrons, where $U$ is the intra-orbital Coulomb repulsion, $U'$ the inter-orbital one, $J$ the Hund's coupling, and $J'$ the pair hopping interaction. Figure 2 depicts the wave-vector dependence of the magnetic fluctuations~\cite{supple}. We find that the most dominant fluctuations are located at ${\bm Q}=(0, 0, 1)$ and $(1, 0, 0)$. The ${\bm Q}$ vector corresponds to the antiparallel alignment of the magnetic moment of two U atoms in the unit cell. This is well consistent with the observed dispersive magnetic excitations by inelastic neutron scattering measurements~\cite{Aeppli87PRL,Goldman}. 
On the other hand, the presence of the sub-dominant peaks at ${\bm Q}=(0, 0, 1/2)$ and $(1, 0, 1/2)$ may correspond to the fragile magnetic phase transition at $T_N\simeq 5$ K [\onlinecite{Aeppli88PRL,Frings,Hayden}]. Indeed, this sub-dominant fluctuation is much enhanced within random phase approximation. However, it needs further investigations along with a problem of magnetic anisotropy. 
Similarly to the previous study~\cite{Nomoto}, the magnetic anisotropy of the uniform susceptibility is not so large, and slightly Ising-type, $\chi_\parallel(0) \gtrsim \chi_\perp(0)$. Although this is the opposite to the experimental observation, we need to consider the large contribution from the localized f-electron part due to the strong electron correlations in the heavy fermion systems. This is a challenging issue in the future, and beyond the scope of this letter.
\begin{figure}[b]
\centering
\includegraphics[width=8.0cm,clip]{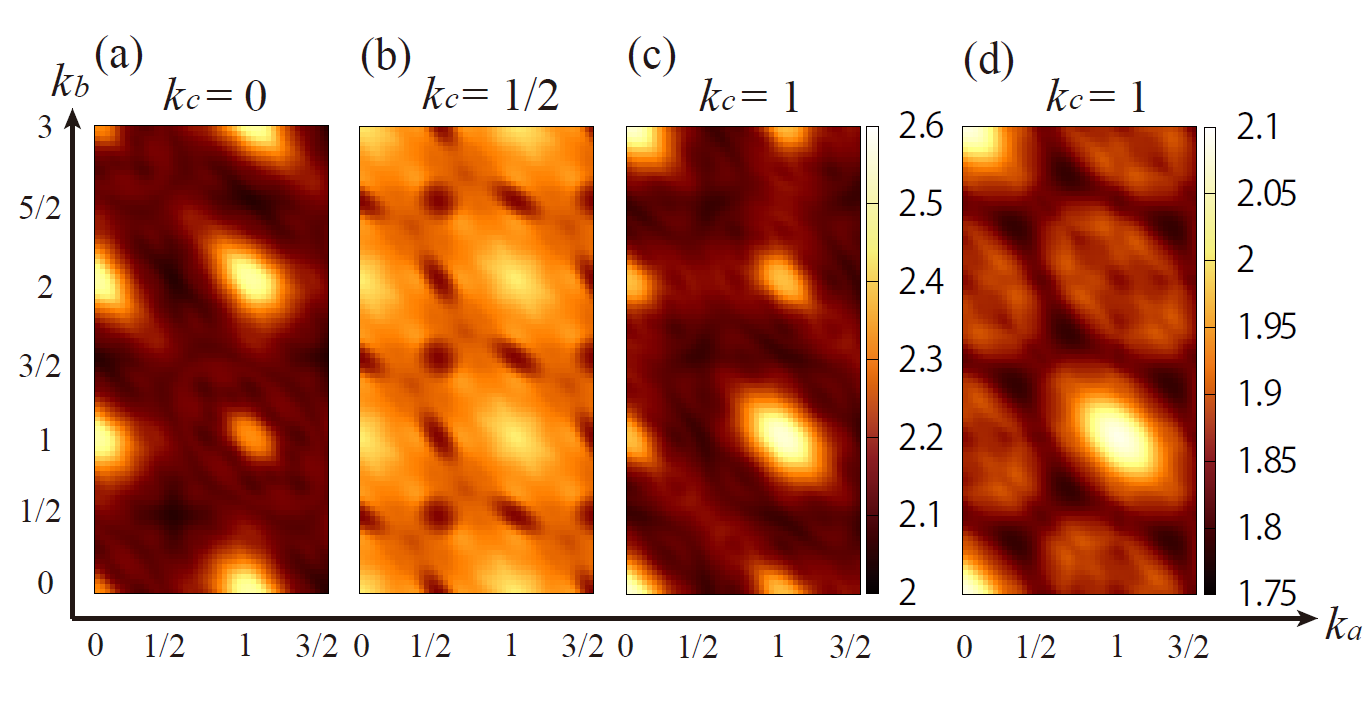}
\caption{Magnetic structure of the bare susceptibilities. (a)-(c) show the magnetic  susceptibilities parallel to $c$-axis, $\chi_\parallel(k_a,k_b,k_c)$, in $k_c=0,~1/2$, and $1$ plane. (d) shows the magnetic susceptibility perpendicular to $c$-axis, $\chi_\perp(k_a,k_b,k_c)$, in $k_c=1$ plane. Difference between (c) and (d) corresponds to the magnetic anisotropy. Note that in actual, the angle between $k_a$ and $k_b$ axes is $\pi/3$.}
\end{figure}

\paragraph*{Superconductivity~---}
\begin{figure}[t]
\centering
\includegraphics[width=6cm]{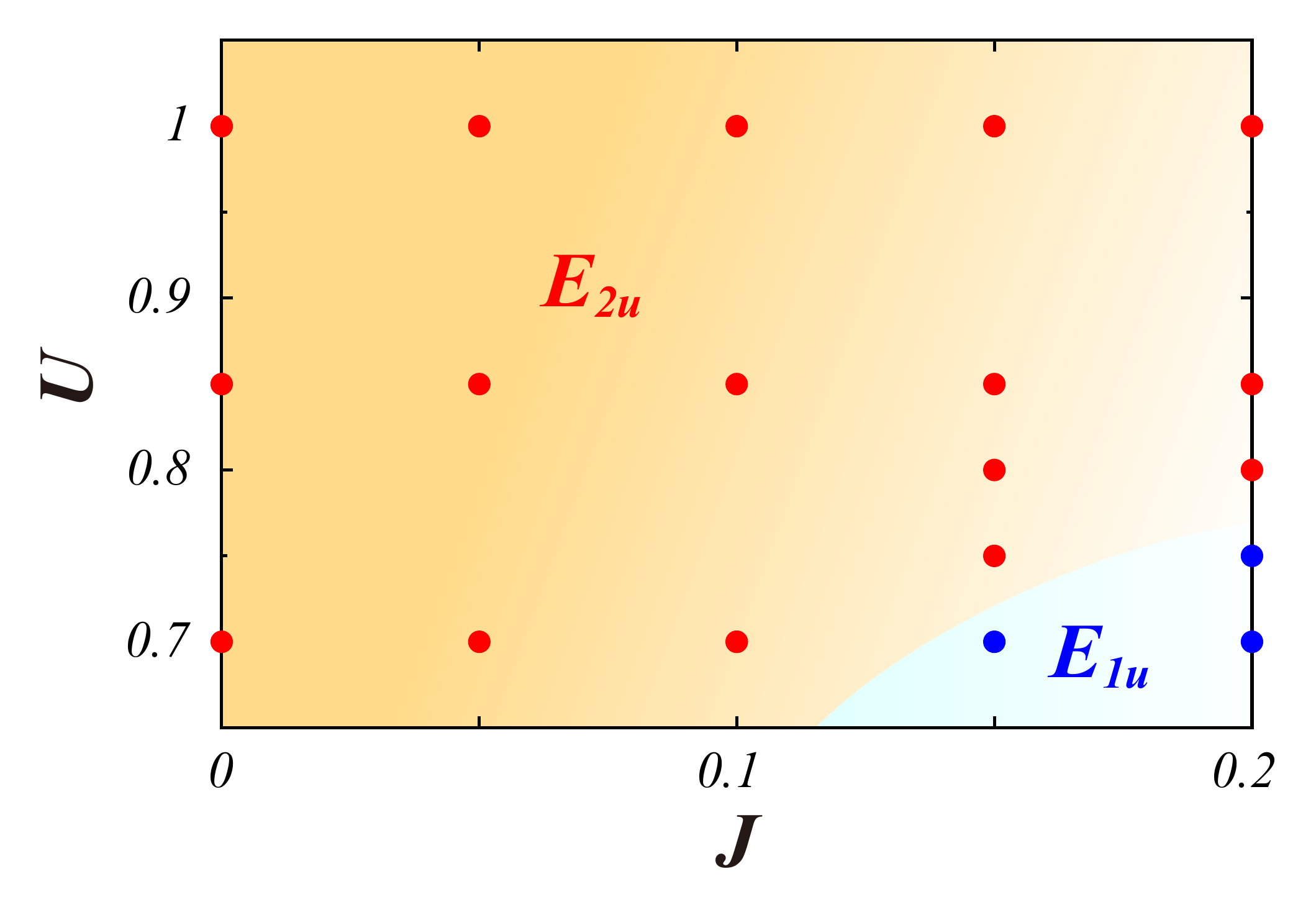}
\caption{Superconducting phase diagram for the intra-orbital on-site repulsion $U$ and Hund's coupling $J$. The unit of energy is eV [\onlinecite{note}]. Here we set the inter-orbital interaction $U'=U$ and the pair hopping $J'=J$. $E_{2u}$ state is predominant over the wide range. Even if assuming $SU(2)$ condition, $U=U'+2J$, the tendency is almost unchanged.}
\end{figure}

Now, let us proceed to a study of the superconducting gap structure. Possible candidates can be obtained by calculating the linearized gap equation at around $T_c$.
\begin{align}
\begin{split}
\lambda \Delta_{\ell m}(k)=\sum_{k'}\sum_{\ell'\ell''m'm''} V_{\ell\ell',m'm}(k-k')~~~~~~~~~~ \\
{\cal G}_{\ell'\ell''}(k'){\cal G}_{m'm''}(-k') \Delta_{\ell''m''}(k'),
\end{split}
\end{align}
where $\Delta_{\ell m}(k)$ and ${\cal G}_{\ell m}(k)$ are the $j$-based gap functions and one-particle Green's functions, and also $\ell$ and $m$ denote $j_z$ components of each U atom~\cite{supple}. The maximum eigenvalue $\lambda$ equals to $1$ at $T_c$.
Here, the pairing interaction $V_{\ell\ell',m'm}(k-k')$ is estimated within the second-order perturbation theory~\cite{note}. It leads to an asymptotically exact weak-coupling solution. In this case, as shown in Fig.3, we obtain two type of predominant spin-triplet pairing states with two dimensional representation $E_{1u}$ and $E_{2u}$. This means that the present microscopic theory supports the phenomenological candidates. In our calculations, the $E_{2u}$ state is more dominant than the $E_{1u}$ state over a wide parameter range. From these results, we conclude that the most promising candidate for the pairing state of UPt$_3$ is the $E_{2u}$ odd-parity state.

Next, let us elucidate the detailed microscopic structure of these pairing states. In Fig.4, we show the superconducting gap amplitude~\cite{supple} on each FS of band$1$, $3$ and $4$. Deep blue corresponds to the gap nodes and/or minima. Slight fluctuation of colors is attributed to the exemplification of the Blount's theorem~\cite{Blount} and some numerical errors. Strictly speaking, the Blount's theorem says that the symmetry-protected line nodes cannot exist in odd-parity representation except for a rare case as discussed later. Therefore, when we do not single out a specific basis function as in the present calculations, the line nodes appear just as a ``pseudo'' line nodes, where the gap amplitude is not exact zero. Hereafter, we  call the ``pseudo'' line nodes by the line nodes. 

It is instructive to start with the $E_{1u}$ state. In such two-dimensional representation, there are two kinds of basis functions. Illustrated in Figs.4(a)-(c) is one possible gap structure in the $E_{1u}$ state. Another one is not shown here. Roughly speaking, the nodal structure on the FS at around the $\Gamma$ point in Fig.4(b) is the f-wave pairing state having one vertical line nodes and two horizontal line nodes at $k_z\neq 0$ plane. This nodal structure is identical to the $E_{1u}$ model, which has been proposed based on the observations in the field-angle resolved thermal conductivity. Since the relevant FS has a large DOS, the in-plane twofold oscillation should be detected also in any experimental observations. However, this is incompatible with the observation in the field-angle resolved specific heat measurement~\cite{Kittaka}.

\begin{figure}[b]
\centering
\includegraphics[width=8.5cm]{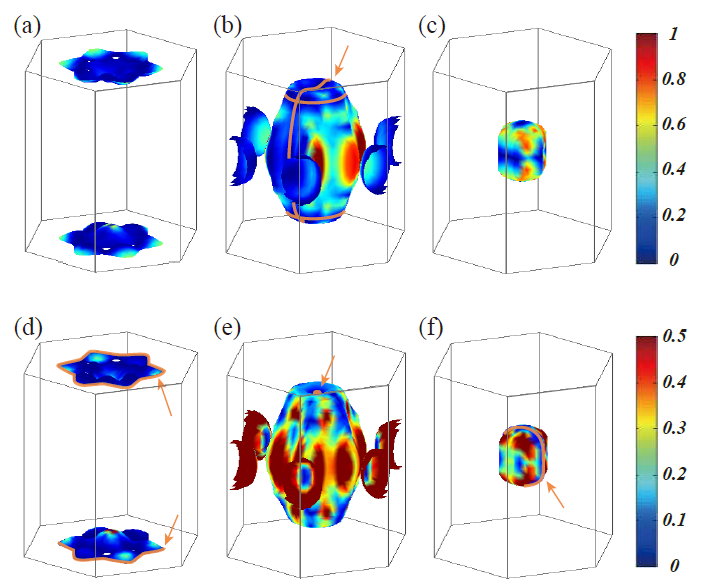}
\caption{Superconducting gap amplitude, $\sum_{n'=\pm n}|\bar{\Delta}_{nn'}(\bk)|^2$, on the FSs of band$1$, band$3$ and band$4$~[\onlinecite{supple}], where $\bar{\Delta}_{nn'}(\bk)=\sum_{\ell m}u_{\ell n}^*(\bk)\Delta_{\ell m}u_{m n'}^*(-\bk)$ with the unitary matrix $u_{\ell n}(\bk)$ diagonalizing $H_0$. $n'=\pm n$ means a sum of the Kramers degeneracy for band$n$. (a)-(c) correspond to the $E_{1u}$ state, and (d)-(f) the $E_{2u}$ state. Line/point nodes colored by orange are pointed by arrows. We recognize that the nodal structure is completely different for each Fermi surface.}
\end{figure}

Furthermore, let us consider the gap structure in the $E_{2u}$ state in Figs.4(d)-(f). Surprisingly, we find that the nodal feature is completely different on each FS; a horizontal nodes in Fig.4(d), point nodes at the top of FS in Fig.4(e), and in-plane twofold vertical line nodes in Fig.4(f). These nodal structures are completely different from those of the previous phenomenological $E_{2u}$ models despite the same irreducible representation. 

Generally, the superconducting order parameter is classified by the irreducible representations of the symmetry in the space group, since the linearized gap equation is separable for each representation, by virtue of the identity property of the pairing interactions. For the strong SOC, symmetry operations act on all the spin, orbital and wave-vector degrees of freedom in our case. If we as usual consider a spin one-half Fermion system without any other internal degrees of freedom, then following Refs.~[\onlinecite{Joynt}] and [\onlinecite{sigrist}], we can see that the only a possible type of $p$-wave gap function in $E_{2u}$ representation is ($\hat{d}_xk_x-\hat{d}_yk_y$, $-\hat{d}_xk_y-\hat{d}_yk_x$) in the ${\bm d}$-vector notation. This minimal gap function has only a point node at the top of FS. Even if considering its higher-harmonics, there does not appear any twofold vertical line nodes. Therefore, it has been widely believed that in $D_{6h}$ point group, twofold vertical line nodes are allowed only in $E_{1u}$ representation, and generally forbidden in $E_{2u}$ representation according to the group theoretical argument~\cite{sigrist}. In this regard, our $E_{2u}$ gap structure seems to be very curious. However, in our case, we need to consider the Cooper pairs in the effective $j=5/2$ space~\cite{Nomoto2}, instead of conventional pseudo-spin $1/2$. Such extension can be performed with the help of projection operator method as in the case of spin $1/2$. Thereby, we find that for the minimal $p$-wave pairing, one of two bases in $E_{2u}$ representation can be described as follows,
\begin{widetext}
\begin{align*}
\Delta_1(\bk)=
\bordermatrix{
& j_z=5/2 & 3/2 & 1/2 & -1/2 & -3/2 & -5/2 \\
& c_1(k_x-ik_y) & c_2 k_z & c_3 k_x+c_4 ik_y & c_5 k_z & c_6 (k_x + ik_y) & 0 \\
& c_2 k_z & c_7 k_x+c_8 ik_y & c_9 k_z & c_{10}(k_x+ik_y) & 0 & c_6 (-k_x+ik_y) \\
& c_3 k_x+c_4 ik_y & c_9 k_z & c_{11}(k_x+ik_y) & 0 & c_{10}(-k_x+ik_y) & c_5 k_z \\
& c_5 k_z & c_{10}(k_x+ik_y) & 0 & c_{11}(-k_x+ik_y) & c_9 k_z & -c_3 k_x+c_4 ik_y \\
& c_6 (k_x+ik_y) & 0 & c_{10}(-k_x+ik_y) & c_9 k_z & -c_7 k_x+c_8 ik_y & c_2 k_z \\
& 0 & c_6 (-k_x+ik_y) & c_5 k_z & -c_3 k_x+c_4 ik_y & c_2 k_z & c_1 (-k_x-ik_y)
},
\end{align*}
\end{widetext}

\noindent
where $c_i~ (i=1\sim 11)$ are material-dependent parameters. From the expressions of the second and fifth diagonal elements, we can verify that twofold vertical line nodes appear in the $j_z=\pm3/2$ subspace. Similarly, we find that the gap functions in the $j_z=\pm5/2$ or $\pm1/2$ subspace yield only point nodes with the linear dispersion along c-axis, and the twofold vertical line nodes are forbidden. Anomalous twofold vertical line nodes in the $E_{2u}$ representation emerge only in the $j_z=\pm 3/2$ space. In UPt$_3$, the FSs in Figs.1(d) and (f) involve plenty of $j_z=\pm 3/2$ component. Thus, it is natural that twofold vertical line nodes emerge in these FSs even in $E_{2u}$ gap symmetry. Moreover, it should be noted that these FSs have a light band mass. In this case, it can be expected that the in-plane twofold oscillation in the field-angle resolved measurements is more prominent in the thermal conductivity than in the specific heat measurements. This can provide an explanation for the seemingly inconsistent observations between these measurements. In addition, since the FS around $\Gamma$ in Fig.1(c) is almost composed of $j_z=\pm5/2$, we recognize that the point nodes observed in Fig.4(e) have linear dispersion, which can be consistent with the small residual thermal conductivity~\cite{Izawa}. 

In order to understand more about this unprecedented $E_{2u}$ gap structure, let us dissect the superconducting gap structure in Fig.4(f). Although the $p$-wave line nodes on the $k_x=0$ plane are remarkable as mentioned above, we can realize additional gap minima on the $k_y=0$ and $k_z=0$ planes. This implies a mixing of $f$-wave component with the form of $k_xk_yk_z\hat{d}_z$, which is indeed allowed in the group-theoretical arguments. Therefore, roughly speaking, the gap structure in Fig.4(f) can be described as a linear combination between the $p$-wave $k_x\hat{d}_x$ and $f$-wave $k_xk_yk_z\hat{d}_z$ in the $\bm{d}$-vector representation in the $j_z=\pm 3/2$ space. Interestingly, in this case, under the applied field parallel to the $c$-axis, the Pauli-limiting behavior will be expected in the upper critical field. Although such suppression has been observed experimentally, we need further investigations, considering the magnetic anisotropy.

Finally, let us comment on the horizontal line nodes at $k_z=\pm 1$ in Fig.4(d). As mentioned above, in an ordinary case, there are only point nodes in $E_{2u}$ representation. However, in the non-symmorphic system like UPt$_3$, there exist additional $C_2$ screw symmetry, which protects the horizontal line nodes. The symmetry-protected line nodes are known as one of exceptions to the Blount's theorem~\cite{Norman}. In the actual situation, however, the interesting line nodes are simply lifted, or slightly shifted from the $k_z=\pm 1$ plane, due to the presence of a weak symmetry-breaking term~\cite{crystal,Hayden}. This is a challenge for the future.

\paragraph*{Conclusion~---}
Based on the advanced first-principles theoretical approach, we clarify the microscopic gap structure in the heavy-fermion superconductor UPt$_3$. We find that the obtained antiferromagnetic fluctuations with ${\bm Q}=(0, 0, 1)$ and $(1, 0, 0)$, which are consistent with the neutron scattering measurements, lead to the spin-triplet pairing states with $E_{1u}$ and $E_{2u}$ representations in the $D_{6h}$ space group. The obtained $E_{1u}$ gap structure is consistent with the phenomenological $f$-wave pairing state. On the other hand, the latter $E_{2u}$ state, having nodal structure different for each band, is distinct from the well-known $E_{2u}$ models. In particular, the in-plane twofold vertical line nodes emerge on the small FS, which can consistently explain the field-angle resolved measurements in both the thermal conductivity and the specific heat. Such peculiar feature cannot be explained in the conventional pseudo-spin representation, but is described by the group-theoretical representation of the Cooper pairs in the $j=5/2$ space. Furthermore, the study of magnetic anisotropy and the mixture of $p$-wave and $f$-wave with different ${\bm d}$-vectors can provide a clue to understand the remaining problems of the Pauli limiting of the upper critical field~\cite{Shivaram, Choi2, Kittaka} and the anomalous behavior of the Knight shift~\cite{Tou} and so on. These are interesting issues in future, together with the understanding of the multiple superconducting phases. Thus, our findings shed new light on the long-standing problems in the superconductivity of UPt$_3$.

\begin{acknowledgements}
We acknowledge Y. Yanase, K. Machida and K. Hattori for valuable discussions, and K. Izawa and Y. Machida for their recent data and helpful discussions. This work was partially supported by JSPS KAKENHI Grant No.15H05745, 15H02014, and 15J01476.
\end{acknowledgements}

\pagebreak
\widetext
\begin{center}
\vspace{1cm}
\textbf{\large Supplemental Materials: \\ \vspace{3mm} Exotic multi-gap structure in UPt$_3$ unveiled by the first-priniciples analysis}
\end{center}
\setcounter{equation}{0}
\setcounter{figure}{0}
\setcounter{table}{0}
\setcounter{page}{1}
\makeatletter
\renewcommand{\theequation}{S\arabic{equation}}
\renewcommand{\thefigure}{S\arabic{figure}}
\renewcommand{\bibnumfmt}[1]{[S#1]}
\renewcommand{\citenumfont}[1]{S#1}

\newcommand{\bra}{\langle \hspace{0.3mm}}
\newcommand{\ket}{\hspace{0.3mm} \rangle}

\section{band structure calculation and effective Hamiltonian}
First of all, we perform the {\it ab initio} band structure calculation in the paramagnetic state of UPt$_3$ using the WIEN2K package~\cite{Wien2kS}, in which the relativistic full-potential (linearized) augmented planewave (FLAPW) + local orbitals method is implemented. The crystallographical parameters are the space group $P6_3/mmc$ which holds the in-plane six-fold rotational symmetry and the experimental lattice constants $a = 5.764$ \AA , $c = 4.899$ \AA \cite{JoyntS}. For the self-consistent calculations, we used PBE-GGA exchange-correlation potential\cite{PerdewS}, $12\times12 \times12$ k-point grid in the Brillouin zone, and a cut-off parameter $RK_{max}=13$. 
The spin-orbit interactions is included with the fully-relativistic calculations.

\begin{figure}[b]
\centering
\includegraphics[width=12.0cm]{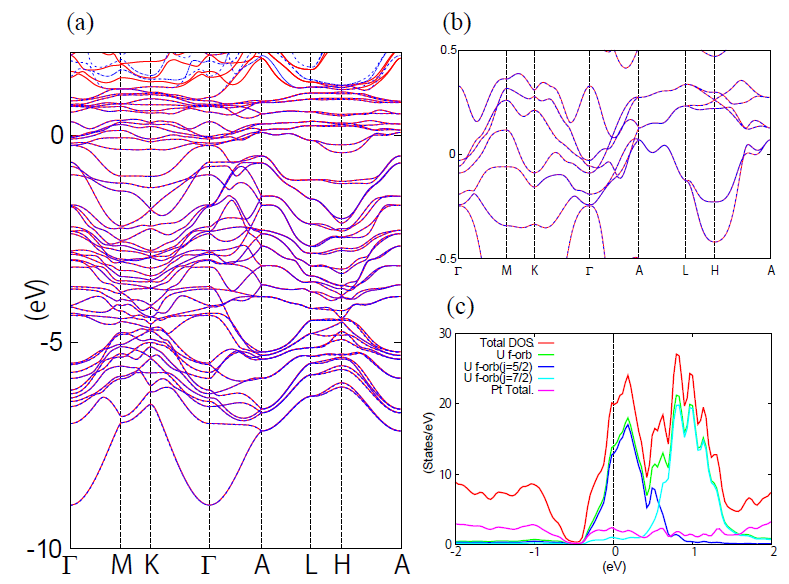}
\caption{(a) Band structure along high-symmetry line and (b) enlarged one near the Fermi level. Red line is the result of {\it ab initio} calculation by WIEN2K. Blue dashed line is the Wannier fit. The dispersion below 1 eV is reproduced completely (c) The partial density of states near Fermi level. }
\end{figure}

In Figure S1, we illustrate the result of the band-structure calculations. Figure S1(a) depicts the GGA band structure (red line) and its Wannier fit~\cite{Wannier90S,KunesS,IkedaS} (blue line). Figure S1(b) is the enlarged figure near the Fermi level. We can see that the fitting works well. The effective tight-binding model is described by $120$ basis functions of U(5f), U(6d), Pt(5d), and Pt(6s) orbitals. The Wannier center of each orbital is located at around the atomic center within numerical errors $10^{-3}$. Figure S1(c) shows the partial density of states (DOS). Blue, cyan, and magenta lines correspond to $j=5/2$ and $7/2$ partial DOS of U(5f), and the total DOS of Pt atoms. We can see that the states crossing the Fermi level are dominated by the U(5f) orbitals, especially, the $j=5/2$ component. The $j=7/2$ states are located around $1$eV higher due to the moderate spin-orbit coupling in U atoms. Therefore, we can expect that the low-energy excitations in this system are dominated by the $j=5/2$ components. Hereafter, we focus on the $j=5/2$ electrons, and regards the other electrons as the conduction electrons. The kinetic term of the effective multi-orbital Anderson model is described as follows,
\begin{align}
H_0=\sum_{\bm k}\biggl\{ \sum_{lm}^{f}\varepsilon_{f \bm k}^{lm}f_{{\bm k}l}^\dagger f_{{\bm k}m}+\sum_{lm}^{c}\varepsilon_{c \bm k}^{lm}c_{{\bm k}l}^\dagger c_{{\bm k}m} \sum_{l}^{f}\sum_{m}^{c} \left(V_{\bm k}^{lm}f_{{\bm k}l}^\dagger c_{{\bm k}m} + V_{\bm k}^{*lm}c_{{\bm k}m}^\dagger f_{{\bm k}l}\right)\biggr\},
\end{align}
where $f^\dagger_l (f_l)$ and $c^\dagger_l (c_l)$ correspond to the creation (annihilation) operators of U(5f) electrons with $j=5/2$ and the other electrons, respectively. The indices $l$ and $m$ run over all of spins, orbitals, and atomic sites degrees of freedom. Moreover, the on-site interactions between $j=5/2$ electrons can be obtained by transformed the following LS-based form into the representation of the $j=5/2$ space,
\begin{align}
H_{\rm int} = \frac{1}{4}\sum_{\bx_i \alpha}\sum_{\zeta_1\zeta_2\zeta_3\zeta_4}^{f}\Gamma^{(0)}_{\zeta_1\zeta_4,\zeta_3\zeta_2}f_{\bx_i \alpha \zeta_1}^\dagger f_{\bx_i \alpha \zeta_2}^\dagger f_{\bx_i \alpha \zeta_3} f_{\bx_i \alpha \zeta_4}, \label{eq:H_int}
\end{align}
where $\zeta_i$ denote both the orbital (angular) and spin quantum numbers. $\bx_i$ denotes a lattice vector and $\alpha$ is a label of two U atoms in a unit cell. $\hat{\Gamma}^{(0)}$ is the Hubbard-type interaction, given by $\Gamma^{(0)}_{\zeta_1\zeta_4,\zeta_3\zeta_2}=-\frac{1}{2}S^{(0)}_{\xi_1\xi_4,\xi_3\xi_2}{\bm \sigma}_{\sigma_1\sigma_4}\cdot{\bm \sigma}_{\sigma_2\sigma_3}+\frac{1}{2}C^{(0)}_{\xi_1\xi_4,\xi_3\xi_2}\delta_{\sigma_1\sigma_4} \delta_{\sigma_2\sigma_3}$ where $\xi_i$ and $\sigma_i$ denote the orbital and spin quantum number respectively. The explicit forms of $\hat{S}^{(0)}$ and $\hat{C}^{(0)}$ are given by,
\begin{align}
\hat{S}^{(0)}=\left\{\begin{aligned}
& U \\
& U' \\
& J \\
& J'
\end{aligned}
\right. ,\hspace{1cm}
\hat{C}^{(0)}=\left\{\begin{aligned}
& U &(\xi_1=\xi_2=\xi_3=\xi_4) \\
& 2J -U' &(\xi_1=\xi_3 \neq \xi_2=\xi_4) \\
& 2U'-J &(\xi_1=\xi_4 \neq \xi_2=\xi_3) \\
& J' &(\xi_1=\xi_2 \neq \xi_3=\xi_4) \\
\end{aligned}
\right.
\end{align}
where $U$ ($U'$) is the intra-orbital (inter-orbital) direct Coulomb interaction, and $J$ and $J'$ represent the Hund's coupling and the pair-hopping interaction. In the actual calculations, we transform this representation of Eq.(\ref{eq:H_int}) into that of $J$ basis, and then neglect the interactions with $j=7/2$ space, since we focus on the low-energy excitations of the $j=5/2$ electrons as mentioned above. As demonstrated in Ref.~[\onlinecite{IkedaS}], momentum dependence of susceptibilities, which is important in unconventional superconductivity, is well described within this approximation.

\section{Green functions and susceptibilities}
Here we provide a brief summary of calculations of susceptibilities within the random phase approximation (RPA)~\cite{IkedaS,NomotoS}. First, the non-interacting Green's functions are given by
\begin{align}
\mathcal{G}_{lm}(k) &=- \int_0^\beta d\tau e^{i\omega_n\tau}\bra T_\tau(f_{\bk l}(\tau) f^\dagger_{\bk m}) \ket_{0}, \\
& = \sum_{m'} \frac{u_{lm'}^\bk u^{\bk\dagger}_{m'm}}{i\omega_n-E_{\bk m'm'}},
\end{align}
where $\hat{u}^{\bk}$ and $\hat{E}_{\bk}$ are the unitary matrix diagonalizing $H_0$ and the energy eigenvalues respectively. The irreducible susceptibilities are defined by
\begin{align}
\chi^{(0)}_{l_1l_4,l_3l_2}(q)=-\frac{T}{N}\sum_{k}\mathcal{G}_{l_1l_3}(k+q)\mathcal{G}_{l_2l_4}(k).
\end{align}
The RPA susceptibility in the matrix form can be obtained as follows,
\begin{align}
\hat{\chi}^{\rm RPA}(q)=(\hat{1}-\hat{\Gamma}^{(0)}\hat{\chi}^{(0)}(q))^{-1}\hat{\chi}^{(0)}.
\end{align}
In general, the magnetic (dipole) correlation functions, $\chi_{ab}({\bm q})$, with $\hat{\chi}(q)=\hat{\chi}^{(0)}(q)$ or $\hat{\chi}^{\rm RPA}(q)$ are given by
\begin{align}
\chi_{ab}({\bm q})&=\sum_{\alpha\alpha'} e^{-i{\bm q}\cdot({\bm \eta}_\alpha-{\bm \eta}_{\alpha'})}\int_0^\beta d\tau \bra T_\tau(J_a^\alpha({\bm q},\tau) J_b^{\alpha'\dagger}({\bm q},0)) \ket\\
&\approx \sum_{\alpha\alpha'} e^{-i{\bm q}\cdot({\bm \eta}_\alpha-{\bm \eta}_{\alpha'})}(J_a^\alpha)_{lm}\chi_{ml,l'm'}({\bm q},0)(J_b^{\alpha'})_{l'm'} \label{eq:supple1},
\end{align}
where ${\bm \eta}_\alpha$ is a position of atom $\alpha$ relative to the lattice vectors and $a,b=x,y$, or $z$. 
The magnetic fluctuation parallel (perpendicular) to the $c$-axis $\chi_\parallel({\bm q})$ ($\chi_\perp({\bm q})$) in the main text is defined by $\chi_\parallel({\bm q})=\chi_{zz}({\bm q})$ ($\chi_\perp({\bm q})=(\chi_{xx}({\bm q})+\chi_{yy}({\bm q}))/2$), given that a total magnetic moment $\hat{M}_a=\hat{L}_a+2\hat{S}_a \simeq g \hat{J}_a$ with the Lande $g$-factor $g = 6/7$. The matrix elements of $\hat{J}_a^\alpha$ can be obtained by the operator equivalent method as usual. From Eq.\eqref{eq:supple1}, the periodicity of $\chi_{ab}({\bm q})$ in UPt$_3$ is $(3,3,2)$ in the unit of reciprocal lattice vector. 
We find that $\chi^{\rm RPA}(q)$ shows some peak structure at ${\bm Q}=(0,0,1)$, $(1,0,0)$, $(0,0,1/2)$, and $(1,0,1/2)$, as discussed for $\chi^{(0)}(q)$ in the main text. For sufficiently large interaction parameters, magnetic fluctuations at the latter two $\bm Q$ vectors are enhanced, and those at the former two are concealed. On the other hand, the superconducting gap structure is not drastically changed for interaction parameters, irrespective of whether the pairing interaction $V_{ll',mm'}(q)$ is given by the second-order perturbation or RPA. 
Based on these results, we restricted ourselves to the weak-coupling approach for simplicity.

\section{Gap functions and classification}
Figure S2 represents all of $E_{1u}/E_{2u}$ gap structure discussed in the main text. These were obtained based on the linearized gap equation Eq.(1) in the main text. Here we describe the way to classify the symmetry of the gap function $\Delta_{lm}(\bk)$, which are expressed as a $12\times12$ matrix in the $j=5/2$ space of two U atoms in the unit cell. 

Let us consider a generic space-group operation $g_s=\{p|{\bm a}\} \in G$, where $G$ is the space group $P6_3/mmc$ in our case, $p$ is a point group operation, and $\bm a$ is a translation associated with $p$. A creation operator $f_{\bk l}^\dagger(=f^\dagger_{\bk\alpha\zeta})$ for a wave-vector $\bk$, atom $\alpha$ and $j_z=\zeta$ is transformed by $g_s$ into
\begin{align}
g_s f_{\bk\alpha\zeta}^\dagger g_s^{-1}=e^{-i\bk'\cdot({\bm\eta}'_\alpha-{\bm \eta}_{\alpha'}+{\bm a})}\sum_{\zeta'}f_{\bk' \alpha'\zeta'}^\dagger D^{(5/2)}_{\zeta'\zeta}(p),
\end{align}
where $\bk'=p\bk$, ${\bm\eta}'_\alpha=p{\bm\eta_\alpha}$, $\alpha'=p\alpha$ and $D^{(5/2)}(p)$ is a representation matrix of $p$ in the $j=5/2$ space, which corresponds to an irreducible representation of $SU(2)$ group. 

In the mean-field approximation, the interaction part of the Hamiltonian is given by
\begin{align}
H_{\rm SC}=\sum_{\bk}\sum_{lm}\Delta_{lm}(\bk)f_{\bk l}^\dagger f_{-\bk m}^\dagger+h.c.
\end{align}
In the group-theoretical classification of gap functions, we study the transformation properties of this Hamiltonian. Given that a gap function $\Delta^{\Gamma i}_{lm}(\bk)$ belongs to an irreducible representation $\Gamma$, $H_{\rm SC}^{\Gamma i}$ meets the following relation, 
\begin{subequations}
\begin{align}
g_s H_{\rm SC}^{\Gamma i} g_s^{-1} &= \sum_{\bk}\sum_{lm} \Delta^{\Gamma i}_{lm}(\bk) (g_s f_{\bk l}^\dagger f_{-\bk m}^\dagger g_s^{-1})+ h.c.\\
&=\sum_{\bk}\sum_{lm} \Delta^{\Gamma j}_{lm}(\bk)D_{ji}^{(\Gamma)}(p) f_{\bk l}^\dagger f_{-\bk m}^\dagger+ h.c.,
\end{align}
\end{subequations}
where $i, j$ denotes a basis of $\Gamma$, and $D^{(\Gamma)}(p)$ is a representation matrix of point group symmetry. 
The gap functions in Fig.S2 have been obtained with such classification. 

Finally, let us comment the $E_{2u}$ gap function in Figs.S2(f)-(j). As mentioned in the main text, symmetry-protected horizontal line nodes appear on the $k_z=\pm \pi$ plane in band1 and band2. Point nodes at $k_x=k_y=0$ are observed in all bands, but remarkable in band3. There emerge twofold symmetric vertical-line gap minima in band4 and band5. Such structure is robust, and observed in a wide range of parameters. On the other hand, the gap structure of band2 is complicated and fragile, since the symmetry argument is not applicable.

\begin{figure}[h]
\centering
\includegraphics[width=16cm]{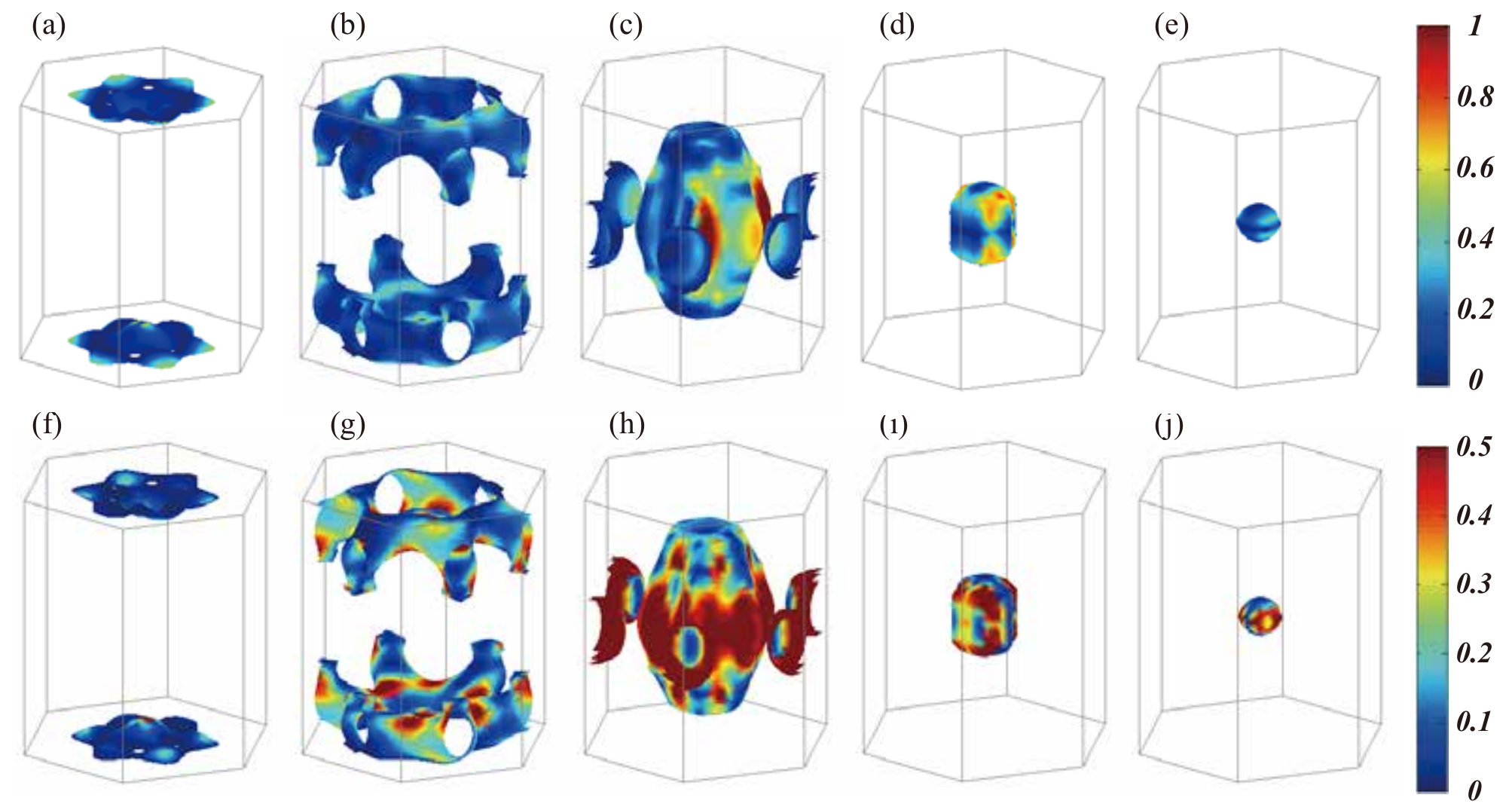}
\caption{Superconducting gap amplitude on all FSs. (a)-(e) correspond to the $E_{1u}$ state, and (f)-(j) the $E_{2u}$ state. Blue color corresponds to the gap nodes and/or minima.}
\end{figure}


\begin{thebibliography}{99}
\bibitem{memo} Unless the Fermi surface in the renormalized band is drastically different from that in the organically GGA band, we can expect that momentum dependence of the pairing interaction and the obtained gap structure will hardly change even considering the correlation effect\cite{yanase}. Hereafter, the correlation effect is considered only through the mass renormalization factor, which is $z \sim 1/20$ from the specific heat coefficient. The interplay between the electron correlation and superconductivity is an interesting issue in future, but it is out of scope in this letter.
\bibitem{note} Based on the Fermi-liquid picture, these interactions should be considered to be quasi-particle interactions, which are renormalized and much smaller than the bare values. The second-order perturbation for the renormalized interactions is valid as asymptotically exact weak-coupling limit. Here we focus on probable spin-triplet states emerging in this limit, although we have used relatively large value of interactions to obtain moderate eigenvalues.
\end{thebibliography}
\end{document}